\documentclass[manuscript]{aastex}

\newcommand{\hst}{{\it HST}}

\newcommand{\ax}{\textrm{[Ar\,{\sc x}]}}
\newcommand{\feii}{Fe~{\sc ii}}
\newcommand{\fex}{\textrm{[Fe\,{\sc x}]}}
\newcommand{\fexiv}{\textrm{[Fe\,{\sc xiv}]}}

\newcommand{\hei}{He~{\sc i}}
\newcommand{\nev}{[Ne~{\sc v}] $\lambda$3426}
\newcommand{\nii}{[N~{\sc ii}]}
\newcommand{\niiip}{\textrm{N\,{\sc iii}}}
\newcommand{\oi}{\textrm{[O\,{\sc i}]}}
\newcommand{\oiii}{[O~{\sc iii}] $\lambda$5007}
\newcommand{\OIII}{[O~{\sc iii}]}
\newcommand{\six}{[\textrm{Si\,{\sc ii}}]}

\newcommand{\kms}{km~s$^{-1}$}
\newcommand{\msun}{{\rm\ M}_\odot}
\newcommand{\rsun}{{\rm\ R}_\odot}
\newcommand{\degree}{$^{\circ}$}

\usepackage{multirow}

\defcitealias{BHO07}{Paper~I}

\shorttitle{Imaging and spectroscopy of RS Oph}
\shortauthors{Ribeiro et al.}

\begin{document}

\title{The Expanding Nebular Remnant of the Recurrent Nova \\
  RS Ophiuchi (2006): II. Modeling of Combined \\
  \emph{Hubble Space Telescope} Imaging and \\
  Ground-based Spectroscopy}

\author{V. A. R. M. Ribeiro,\altaffilmark{1} M. F. Bode,\altaffilmark{1} M. J. Darnley,\altaffilmark{1} D. J. Harman,\altaffilmark{1} A. M. Newsam,\altaffilmark{1} \\ T. J. O'Brien,\altaffilmark{2} J. Bohigas,\altaffilmark{3} J. M. Echevarr\'{i}a,\altaffilmark{4} H. E. Bond,\altaffilmark{5} V. H. Chavushyan,\altaffilmark{6} \\ R. Costero,\altaffilmark{4} R. Coziol,\altaffilmark{7} A. Evans,\altaffilmark{8} S. P. S. Eyres,\altaffilmark{9} J. Le\'on-Tavares,\altaffilmark{6,10} \\ M. G. Richer,\altaffilmark{3} G. Tovmassian,\altaffilmark{3} S. Starrfield,\altaffilmark{11} and S. V. Zharikov\altaffilmark{3}}
\altaffiltext{1}{Astrophysics Research Institute, Liverpool John Moores University, Twelve Quays House, Egerton Wharf, Birkenhead, Wirral, CH41 1LD, UK; (var, mfb, mjd, dh, amn)@astro.livjm.ac.uk}
\altaffiltext{2}{Jodrell Bank Centre for Astrophysics, School of Physics and Astronomy, University of Manchester, Alan Turing Building, Manchester, M13 9PL, UK; tim.obrien@manchester.ac.uk}
\altaffiltext{3}{Instituto de Astronom\'ia, Universidad Nacional Aut\'onoma de M\'exico, Apartado Postal 877, 22830, Ensenada, Baja California, M\'exico; (jbb, richer, gag, zhar)@astrosen.unam.mx}
\altaffiltext{4}{Instituto de Astronom\'ia, Universidad Nacional Aut\'onoma de M\'exico, Apartado Postal 70-264, 04510 M\'exico D.F., M\'exico; (jer, costero)@astroscu.unam.mx}
\altaffiltext{5}{Space Telescope Science Institute, Baltimore, MD 21218, USA; bond@stsci.edu}
\altaffiltext{6}{Instituto Nacional de Astrof\'{i}sica \'{O}ptica y Electr\'onica, Puebla, A.P. 51-216, M\'{e}xico; (vahram, fleon)@inaoep.mx}
\altaffiltext{7}{Departamento de Astronom{\'i}a, Universidad de Guanjuato, A.P. 144 3600 Guanajuato, Gto M\'exico; rcoziol@astro.ugto.mx}
\altaffiltext{8}{Astronomy Group, School of Physical and Geographical Sciences, Keele University, ST5 5BG, UK; ae@astro.keele.ac.uk}
\altaffiltext{9}{Center for Astrophysics, University of Central Lancashire, Preston, PR1 2HE, UK; spseyres@uclan.ac.uk}
\altaffiltext{10}{Max-Planck-Institut f\"ur Radioastronomie, Auf dem H\"ugel 69, 53121, Germany}
\altaffiltext{11}{School of Earth and Space Exploration, Arizona State University, Tempe, AZ 85287-1404, USA; sumner.starrfield@asu.edu}

\begin{abstract}
We report {\it Hubble Space Telescope} imaging, obtained 155 and 449 days after the 2006 outburst of the recurrent nova RS Ophiuchi, together with ground-based spectroscopic observations, obtained from the Observatorio Astron\'{o}mico Nacional en San Pedro M\'{a}rtir, Baja California, M\'exico and at the Observatorio Astrof\'isico Guillermo Haro, at Cananea, Sonora, M\'exico. The observations at the first epoch were used as inputs to model the geometry and kinematic structure of the evolving RS Oph nebular remnant. We find that the modeled remnant comprises two distinct co-aligned bipolar components; a low-velocity, high-density innermost (hour glass) region and a more extended, high-velocity (dumbbell) structure. This overall structure is in agreement with that deduced from radio observations and optical interferometry at earlier epochs. We find that the asymmetry observed in the west lobe is an instrumental effect caused by the profile of the \hst\ filter and hence demonstrate that this lobe is approaching the observer. We then conclude that the system has an inclination to the line of sight of 39$^{+1}_{-10}$ degrees. This is in agreement with the inclination of the binary orbit and lends support to the proposal that this morphology is due to the interaction of the outburst ejecta with either an accretion disk around the central white dwarf and/or a pre-existing red giant wind that is significantly denser in the equatorial regions of the binary than at the poles. The second epoch \hst\ observation was also modeled. However, as no spectra were taken at this epoch, it is more difficult to constrain any model. Nevertheless, we demonstrate that between the two \hst\ epochs the outer dumbbell structure seems to have expanded linearly. For the central (hour glass) region there may be evidence of deceleration, but it is harder to draw firm conclusions in this case.
\end{abstract}


\section{INTRODUCTION}
RS Ophiuchi is a symbiotic recurrent nova (RN) with outbursts recorded in 1898, 1933, 1958, 1967, 1985 \citep[see][]{R87,RI87} and 2006 \citep[][and references therein]{EBO08}. However, during the period 1898 - 1933, before the star was known to be a RN, some outbursts might have occurred, most likely in 1907 \citep{S04} and there is also evidence for an eruption in 1945 \citep{OM93}. RS Oph was observed to be undergoing its latest eruption on 2006 February 12.83 \citep{NHK06}, reaching a magnitude V = 4.5 at this time. We define this as $t$ = 0. The optical light curve then began a rapid decline, consistent with that seen in previous outbursts \citep[][AAVSO\footnote{see http://www.aavso.org}]{R87}. The distance to RS Oph has been derived from several lines of evidence as 1.6 $\pm$ 0.3 kpc (\citealt{B87}; \citealp[see also][]{BMS08} and the discussion in Section 3.1 below).

The RS Oph binary system comprises a red giant, with a spectral type estimated to be in the range G5III to M4III \citep[e.g.][]{BED89,AM99}, and a white dwarf (WD) with mass thought to be close to the Chandrasekhar limit \citep[see][]{DK94,SKS96,FJH00,BQM09}. Accretion of hydrogen-rich material from the red giant onto the WD surface leads to a thermonuclear runaway, as in the outbursts of  classical novae (CNe). The much shorter inter-outburst period for RNe compared to CNe is predicted to be due to a combination of high WD mass and a high accretion rate \citep[e.g.][]{SST85,YPS05}. These models lead to the ejection of smaller amounts of material at higher velocities than those for CNe (typically 10$^{-8}$ to 10$^{-6}$ $\msun$ and several thousand \kms, respectively, for RNe). Spectroscopy of RS Oph has indeed shown H$\alpha$ line emission with FWHM = 3930 \kms\ and FWZI = 7540 \kms\ on 2006 February 14.2 ($t$ = 1.37 days; \citealt{B06}).

Unlike CNe, where the mass donor is a low mass main sequence star, the presence of the red giant in the RS Oph binary means that the high-velocity ejecta run into a dense circumstellar medium in the form of the red giant wind, setting up a shock system. The forward shock moving into the wind has a gas temperature $\sim$ 2.2$\times$10$^8$ K for a shock velocity $v_s$ = 4000 \kms\ \citep{BOO06}.

Very Long Baseline Interferometry (VLBI) radio observations by \citet{OBP06} at t = 13.8 days showed a partial ring of non-thermal radio emission resulting from the expanding shock, which developed into a more bipolar structure \citep[see also][]{OBP08}. They suggested that the asymmetry in this ring could be due to absorption in the overlying red giant wind and they also noted the emergence of more extended components to the east and west. \citet{SRM08} followed the evolution of the radio source with the Very Long Baseline Array (VLBA) at 4 epochs between 34 and 51 days after outburst. They found a central thermally-dominated source linked by what appeared to be a narrow (collimated) outflow to expanding non-thermal lobes which they interpreted as the working surfaces of jets. Jet collimation could be due to the expected accretion disc around the WD. It should be noted that \citet{TDP89} had also interpreted their 5 GHz VLBI map from day 77 after the 1985 outburst in terms of a central thermal source with expanding non-thermal lobes.

From {\it Hubble Space Telescope (\hst)} observations, 155 days after outburst, \citet[][hereafter \citetalias{BHO07}]{BHO07} revealed the remnant to have a double ring structure with the major axis lying in the east-west direction (see also Section \ref{results}). They suggested that there was evidence for deceleration in the north-south direction by comparing their observations to earlier epoch observations in the radio \citep{OBP06}. \citetalias{BHO07} also provided preliminary models of the remnant as a bipolar structure which implied a true expansion velocity, $V_{\textrm{exp}}$ = 5600 $\pm$ 1100 \kms\ for the material at the poles. It was proposed that the shaping of the remnant occurred due to the ejecta interaction with the pre-existing circumstellar environment.

The projected 2-D geometry of nova explosions on the sky is usually complex, as are their 1-D spectra. However, taken together they provide the underlying 3-D structure of an object. The projected image on the sky provides spatial information on its structure while the spectra provide information about the component of the velocity vector along the line of sight. Here we re-examine the \hst\ observations of the expanding nebular remnant taken at $t=$ 155 days and report those at a second epoch at $t=$ 449 days after outburst. We compare these to spectral observations (where available), then using a modeling code, determine the detailed 3-D structure of the remnant. In Section \ref{obs} we present our observations and data reduction methods. The results are presented in Section \ref{results} and then modeled in Section \ref{model}. In Section \ref{discussion} we discuss the implications of the results found in Section \ref{model}. Finally, in Section \ref{future} we summarise our conclusions.

\section{OBSERVATIONS AND DATA REDUCTION}\label{obs}
We imaged RS Oph with the \hst\ on two occasions, under the Director's Discretionary programs GO/DD-11004 and GO/DD-11075. The first observations were made on 2006 July 17 (155 days after outburst) with the High-Resolution Channel (HRC) of the Advanced Camera for Surveys (ACS), with a scale of $0\farcs025$ pixel$^{-1}$ (equivalent to 40 AU pixel$^{-1}$ at $d = 1.6$ kpc).  The second observations were made on 2007 May 7 (449 days after outburst) with the Planetary Camera CCD of the Wide Field Planetary Camera 2 (WFPC2), with a scale of $0\farcs046$ pixel$^{-1}$, since in the meantime the ACS had suffered a failure. During the first epoch observations, the ACS/HRC was used with three narrowband filters to isolate the H$\alpha$+\nii\ (F658N), \oiii\ (F502N) and \nev\ (F344N) nebular emission lines. For the subsequent WFPC2 observations, only the F502N filter was used in order to isolate the \oiii\ emission line. This line was chosen based on the results from the first epoch observations.

Here we concentrate on the F502N filter observations common to both epochs. The other two \hst\ filter bands, F344N and F658N, will not be discussed because we have no spectral information available for the former and the latter presented little significant structure in the epoch 1 \hst\ images \citepalias[see][]{BHO07}.

All data were reprocessed using standard procedures outlined in the ACS\footnote{See http://www.stsci.edu/hst/acs/documents/handbooks/\\DataHandbookv5/ACS\_longdhbcover.html} and WFPC2\footnote{See http://www.stsci.edu/instruments/wfpc2/Wfpc2\_dhb/\\WFPC2\_longdhbcover.html} Data Handbooks and the Pydrizzle\footnote{See http://stsdas.stsci.edu/pydrizzle} and Multidrizzle\footnote{See http://stsdas.stsci.edu/pydrizzle/multidrizzle/} Handbooks. PSF profiles were generated using TinyTim \citep{K95} for both ACS/HRC and WFPC2 images. In the ACS/HRC and WFPC2 images, the PSF flux was scaled to an extent such that when subtracted from the observed image the new PSF-subtracted image pixel flux was close to those surrounding it. Deconvolution was then performed using the Lucy-Richardson method for the ACS/HRC and WFPC2 images. CLEAN and Maximum Entropy techniques were also used as a check, and produced similar results.

Spectral observations were carried out at the Observatorio Astron\'omico Nacional en San Pedro M\'artir, Baja California, M\'exico, using the 2.1 m telescope with the Boller \& Chivens and Echelle spectrographs, and at the Observatorio Astrof\'\i{}sico Guillermo Haro, at Cananea, Sonora, M\'exico, using the 2.12 m telescope with their Boller \& Chivens spectrograph. The Boller \& Chivens spectrographs have intermediate spectral resolutions, from R $\sim$ 1500 to 3500, while the Echelle spectrograph has a maximum resolution of about R = 18000 at 5000\AA. The data were reduced using the Image Reduction and Analysis Facility (IRAF)\footnote{IRAF is distributed by the National Optical Astronomical Observatories, which is operated by the Associated Universities for Research in Astronomy, Inc., under contract to the National Science Foundation.} software.

The observations were carried out over several epochs from 10 days (2006 February 22) to  168 days (2006 July 30) after outburst. The log of spectroscopic observations is shown in Table \ref{tb:spectra_log}. For direct comparison with the \hst\ imaging we focus on the spectral wavelength region  around the \oiii\ emission line, which falls in the \hst\ F502N narrowband filter.

\section{RESULTS OF OBSERVATIONS}\label{results}
\subsection{Hubble Space Telescope Images}\label{results_hst}
\begin{figure}
\plotone{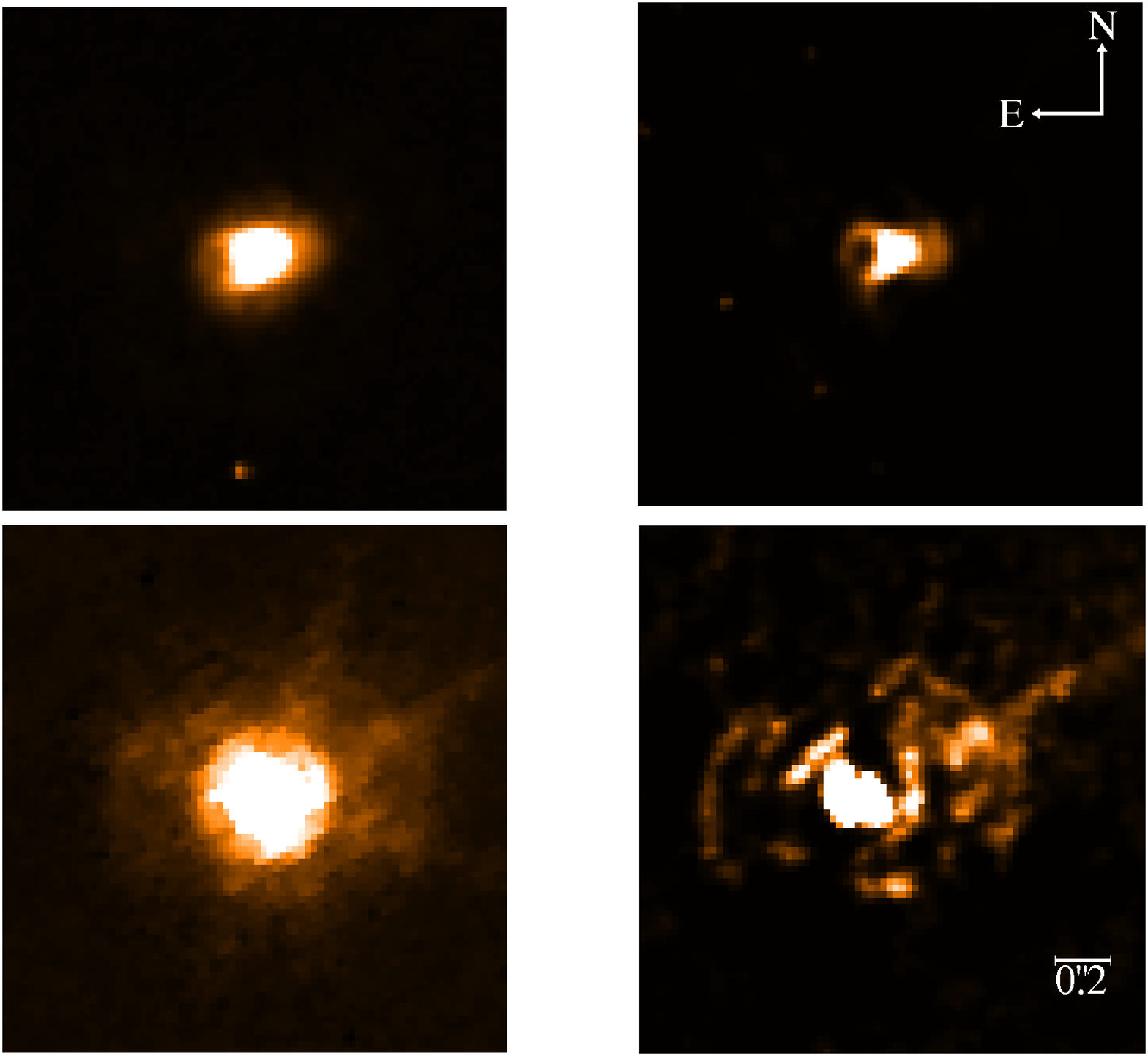}
\caption{\hst\ images of RS Oph through the F502N filter. {\it Top left} - Raw image from the first epoch observations (t = 155 days after outburst) using the ACS/HRC. {\it Top right} - PSF-subtracted and deconvolved image using a TinyTim PSF, showing a double ring structure. {\it Bottom left} - Raw image from the second epoch observations (t = 449 days after outburst) using WFPC2. {\it Bottom right} - PSF-subtracted and deconvolved image using a TinyTim PSF, again showing evidence of a double ring structure (note that the ``jet-like'' feature on the WFPC2 deconvolved image is an artifact caused by bleeding on the CCD chip). North is up and east is to the left in all images.}
\label{fig:hst_obs}
\end{figure}

The first epoch observations clearly show extended structure in the \oiii\ line in the deconvolved images, as well as in the raw images at both epochs (Figure \ref{fig:hst_obs}). As reported in \citetalias{BHO07}, a striking feature is the double ring structure (top image in Figure \ref{fig:hst_obs}) with the major axis lying east-west with a total (peak-to-peak) extent of 360 $\pm$ 30 mas, in the plane of the sky, corresponding to an expansion rate of 1.2 $\pm$ 0.1 mas day$^{-1}$ (and equivalent to $V_{\textrm{t}}$ = 3200 $\pm$ 300 \kms\ for $d = 1.6$ kpc, where $V_{\textrm{t}}$ is the transverse velocity). As an aside, we note that a much larger distance of 4.3 kpc has been suggested by \cite{S09}. If correct, that would result in $V_{\textrm{t}}$ = 8600 \kms\ which is of course very much a lower limit to the true expansion velocity of material and which appears to be unfeasibly large for a nova outburst. We also note that the most recent compilation of all distance estimates by \cite{BMS08} gives $d = 1.4^{+0.6}_{-0.2}$ kpc, in line with the value we assume here. More extended emission is detected above background in the deconvolved image with an expansion rate of 1.7 $\pm$ 0.2 mas day$^{-1}$ and, when compared with values seen in the radio, it was suggested in \citetalias{BHO07} that the east-west bipolar emission seen here and in the radio arises from the same regions of the remnant, if the expansion velocities in the east-west direction are roughly constant after outburst \citep[see also][]{OBP08}.

The north-south (peak-to-peak) extent is 150 $\pm$ 25 mas corresponding to an expansion rate of 0.48 $\pm$ 0.08 mas day$^{-1}$. Results of the expansion in the early radio observations \citep{OBP06,RMS08} and more recently from \citet{OBP08}, led to a north-south expansion rate of 0.77 $\pm$ 0.04 mas day$^{-1}$ derived from observations over the first 107 days, suggesting deceleration in this direction \citepalias{BHO07}.

In the second epoch images (bottom images on Figure \ref{fig:hst_obs}), structure is still visible in the \oiii\ line. It has an east-west extent of 1100 $\pm$ 100 mas (peak-to-peak) corresponding to  an expansion rate from the center of 1.2 $\pm$ 0.1 mas day$^{-1}$ suggesting no deceleration had taken place between epochs in terms of greatest east-west extent. The north-south extent is much harder to determine but is approximately 460 $\pm$ 46 mas corresponding to an expansion rate of 0.51 $\pm$ 0.03 mas day$^{-1}$. This result would imply that no deceleration occurred in the north-south direction between the two epochs.

It is interesting to compare the derived extent of the remnant at each epoch with the expected dimensions of the red giant wind. Assuming a wind that has been expanding for 21 yrs (the period between the last two outbursts) and a wind velocity of 40 \kms\ \citep{W58}, this implies a maximum wind radius of 2.6$\times$10$^{15}$ cm. For a distance $d = 1.6$ kpc, the first epoch angular size suggests the outburst remnant has maximum east-west extent from the center of 4.3 $\pm$ 0.4$\times$10$^{15}$ cm and at the second epoch the remnant has maximum extent 1.3 $\pm$ 0.1$\times$10$^{16}$ cm. Therefore, by the time of the \hst\ observations the remnant appears larger in the east-west direction than the expected size of the red giant wind.

\subsection{Spectroscopy}
\begin{figure}
\plotone{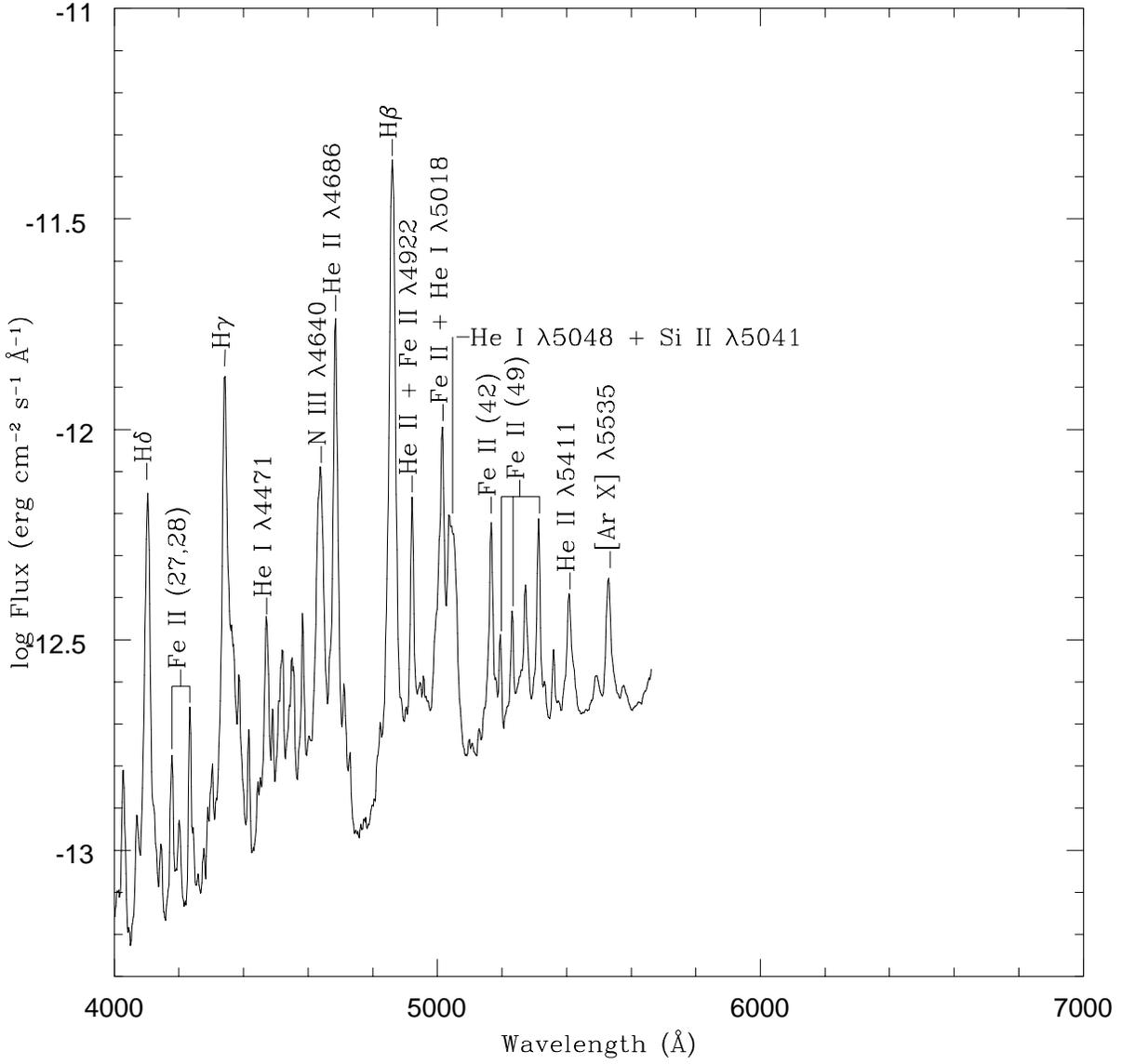}
\caption{Spectral observations of RS Oph on 2006 March 15 (t = 31 days after outburst - see Table \ref{tb:spectra_log}).}
\label{fig:mar_bch}
\end{figure}

\begin{figure}
\plotone{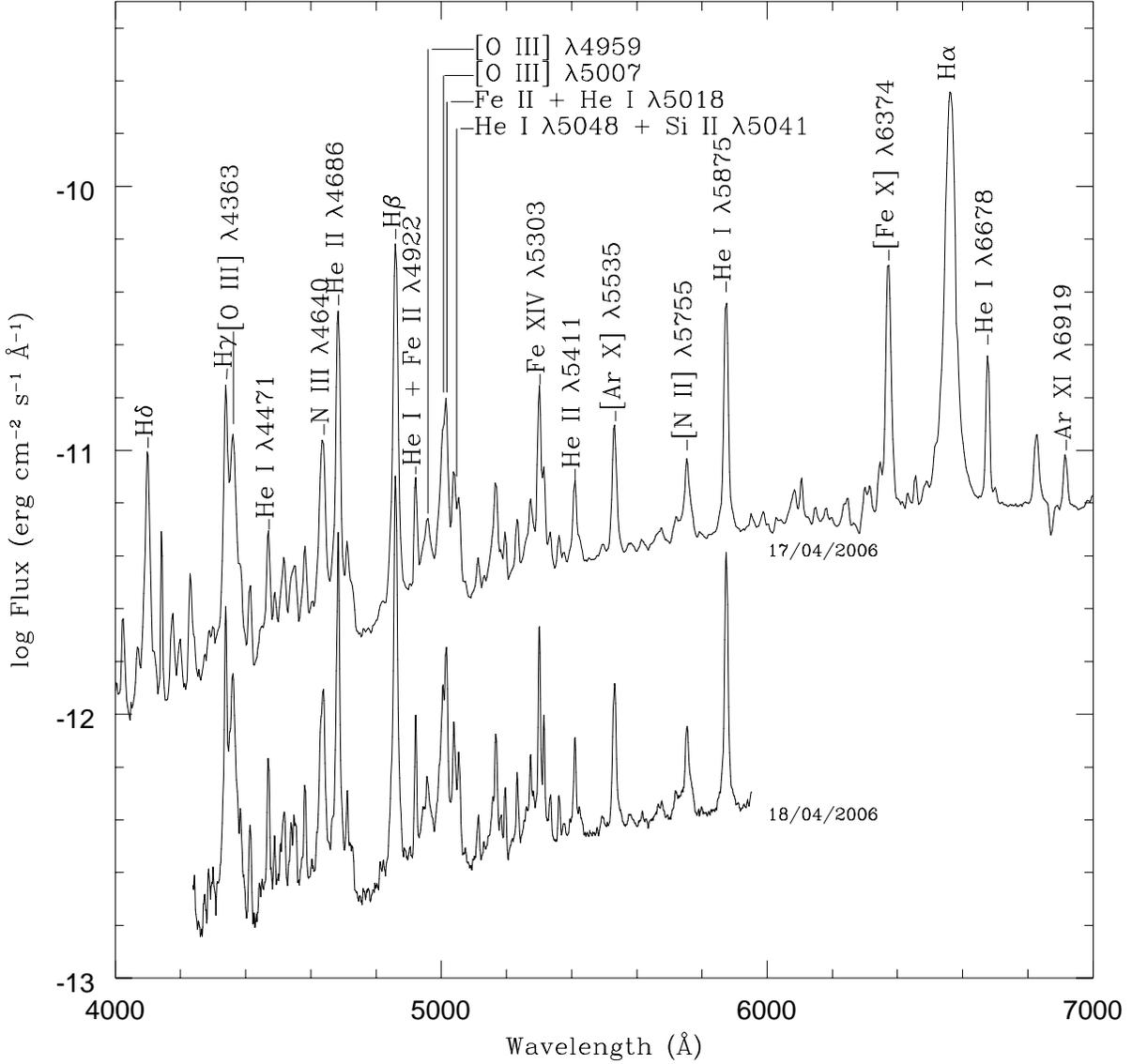}
\caption{As Figure \ref{fig:mar_bch}, but for 2006 April 17 and 18 ($t = 64 - 65$ days after outburst). Note that the day 64 spectrum has been offset in flux by +1 dex for clarity.}
\label{fig:april_bch}
\end{figure}

\begin{figure}
\plotone{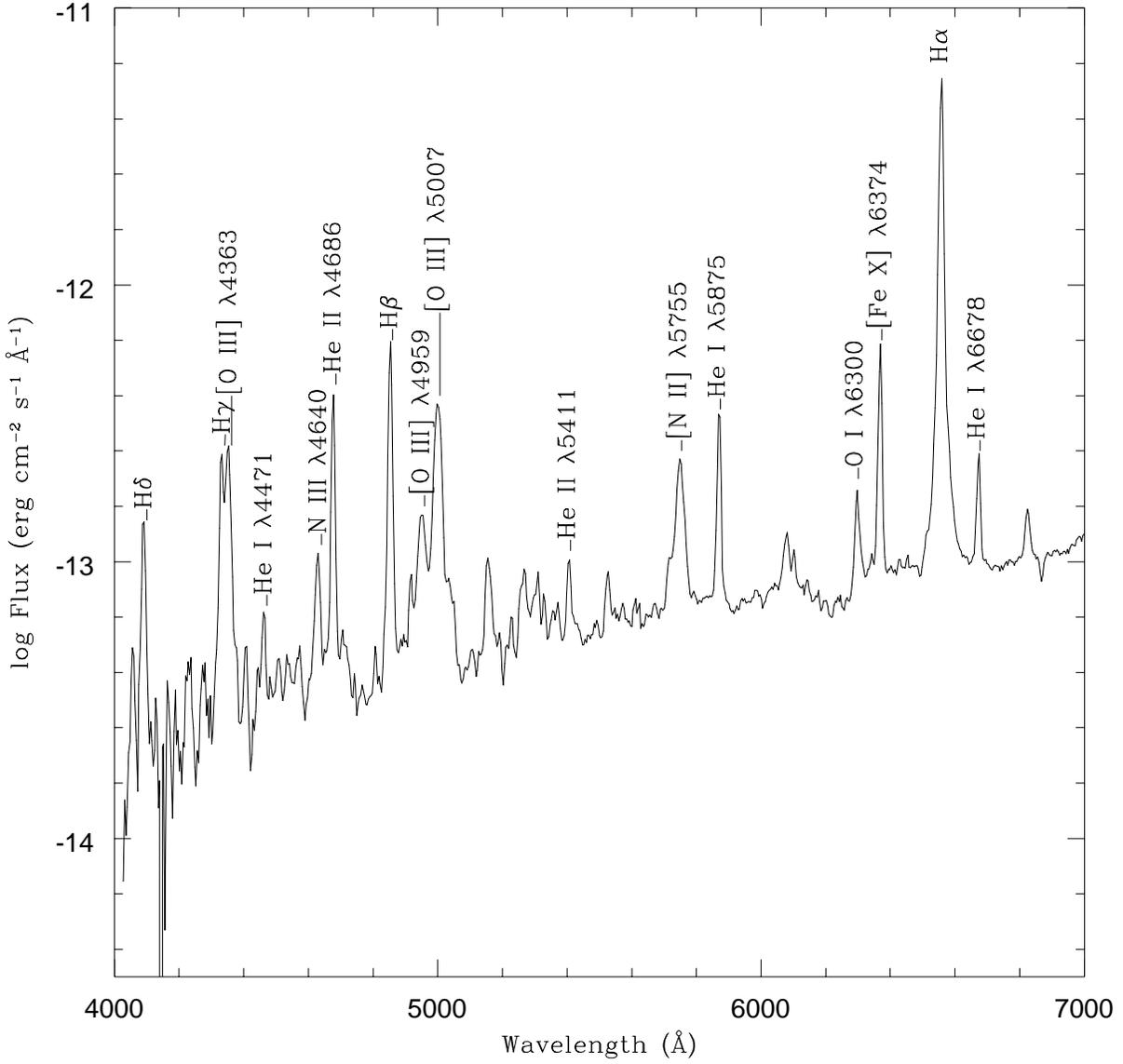}
\caption{As Figure \ref{fig:mar_bch}, but for 2006 May 29 (t = 107 days after outburst).}
\label{fig:may_bch}
\end{figure}

In Figures \ref{fig:mar_bch}-\ref{fig:may_bch} we present the spectral evolution of RS Oph as determined from low resolution spectra obtained from day 31 to day 83 after outburst. The principal emission lines were identified following \citet{R87}. Figure \ref{fig:mar_bch} shows the observed spectrum 31 days after outburst. The visual magnitude had declined to V $\sim$ 9 by this time. The spectrum shows the presence of broad emission lines of H$\beta$, H$\gamma$, H$\delta$, \hei\ ($\lambda$4471, 4713, 4922, 5016, 5048 and 5411), \feii\ (multiplets 27, 28, 42, 48, 49), \niiip\ $\lambda$4640 and \six\ $\lambda$5041. What \citet{R87} identified as the \ax\ $\lambda$5535 coronal line is also present. Figure \ref{fig:april_bch} shows spectra taken on days 64 and 65 after outburst. The magnitude had now declined to V $\sim$ 10. The spectra show an increased degree of excitation, with the presence of moderately strong coronal lines of \ax\ $\lambda$5535 plus \fexiv\ $\lambda$5303, and \fex\ $\lambda$6374. Also present at this time  are the nebular lines of \OIII\ ($\lambda$4363, 4959 and 5007) and \nii\ $\lambda$5755. Figure \ref{fig:may_bch} shows spectra on day 107 after outburst (V $\sim$ 11.5, i.e. approaching minimum light). Here the \fexiv\ $\lambda$5303 and \ax\ $\lambda$5535 lines have become significantly weaker. In contrast, the spectrum also shows an increase in the relative strength of \OIII, and \nii, and \oi\ $\lambda$6300 has appeared. This spectral evolution is typical of that seen in previous outbursts \citep{R87,RI87}. We also note, as an aside to our main discussion, that in the day 64 spectrum for example, the lines identified by \citet{R87} and others as coronal lines are blueshifted with respect to the Balmer and other lower excitation lines by $\sim 100$ km s$^{-1}$. This may be related to the relative principal regions of origin of these lines within the velocity field of the remnant. 

\begin{figure}
\plotone{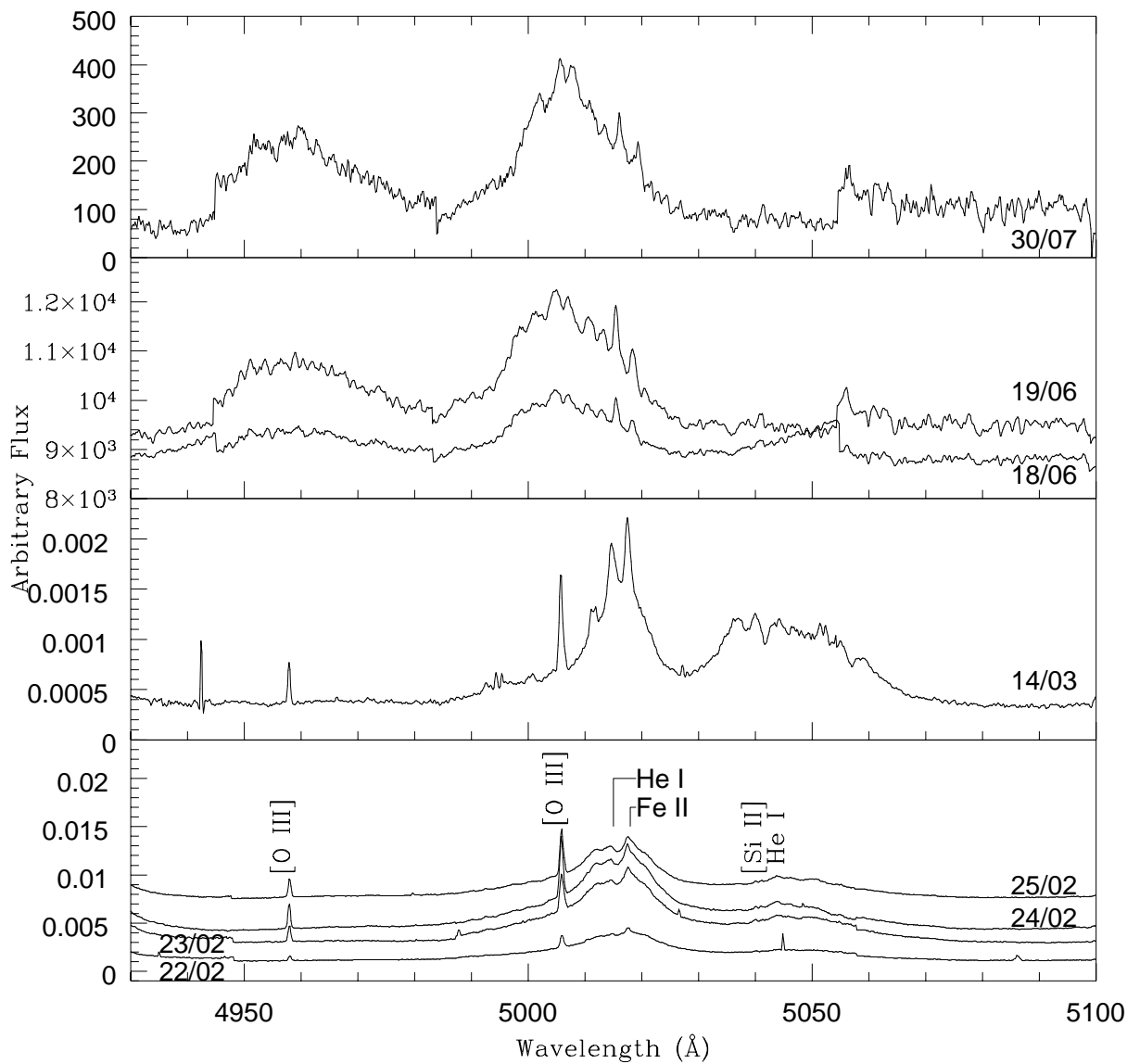}
\caption{RS Oph \oiii\ spectral evolution at different epochs after the 2006 outburst. From bottom to top, 2006 February, March, June and July (dates as shown). Here the evolution of \oiii\ can be tracked first as arising from the ionised wind of the red giant (narrow lines in February and March) and then at later times (June and July) it is more likely related to shocked ejecta and/or shocked wind (broad lines). See Table \ref{tb:spectra_log} for more details of spectroscopic observations.}
\label{fig:evolution}
\end{figure}

We used the higher resolution spectra in the region of the \oiii\ emission line to distinguish features arising from different parts of the remnant (Figure \ref{fig:evolution}). For example, during the months of 2006 February and March (bottom two panels in Figure \ref{fig:evolution}, approximately $\sim$ 10 to $\sim$ 30 days after outburst), we identified emission lines of \OIII\ $\lambda$4959 and $\lambda$5007, \hei\ $\lambda$5016 and $\lambda$5048, \feii\ $\lambda$5018 and \six\ $\lambda$5041. We can immediately see that \hei\ $\lambda$5048 and \six\ $\lambda$5041 lines are not strongly present in the spectra in 2006 June and July (top two panels in Figure \ref{fig:evolution}). Here we interpret narrow emission lines as arising from the ionized red giant wind ahead of the forward shock, whereas the broad features are associated with the shocked ejecta or the shocked wind \citep[see e.g.][]{SKS96}.

With the knowledge of the likely origin of various features in the \oiii\ spectral region, we can start modeling the emission in this line from the expanding remnant alone (i.e. by ignoring emission from other elements or from the ionized wind - see Section \ref{model} below). The line profile we model in detail is that of 2006 July 30 ($t = 168$ days) since it was obtained closest in time to our \hst\ imaging (top panel Figure \ref{fig:evolution}).

\section{MODELING THE REMNANT}\label{model}
Numerical simulations by \citet{LBO93} suggested that bipolarity can be achieved by either ($i$) the outburst being spherically symmetric and the bipolarity resulting from subsequent interactions with an anisotropic red giant wind or ($ii$) the outburst itself being intrinsically bipolar, with material being ejected preferentially in the polar directions and then interacting with an isotropic ambient medium. \citet{LBO93} found the latter to be the best fit to (the admittedly sparse) observations of bipolar structure from VLBI observations of the 1985 outburst of RS Oph.

More recently, \citet{SRM08} suggested that the observed bipolar structure in the radio was not due to an intrinsically asymmetric explosion or to shaping of the ejecta by circumbinary material but due to collimation at very early times, possibly by the accretion disk. Other authors \citep{SPB08,BDA09} have modeled the RS Oph outburst by means of Gaussian fitting to line profiles and find several components that are interpreted as arising from different regions of the expanding remnant; namely the central broad component is associated with more slowly expanding material from the waist while the outliers of the lines are associated by these authors with high velocity ejecta.

Here we have used {\it Shape} \citep{SL06} to analyse and disentangle the 3-D geometry and kinematic structure of the expanding nebular emission in RS Oph. {\it Shape} was originally developed to model the complex structures of Planetary Nebulae and is based on computationally efficient mathematical representations of the visual world which allow for the construction of objects placed at any orientation in a cubic volume. It allows modeling of the structure and kinematics of an object to compare with observed images and spectra and includes parameters such as location and width of the spectrograph slit, seeing values and spectral resolutions. We note here that the nebular remnant was of course much smaller than the width of the spectrograph slit at the time of the contemporaneous {\em HST} observations on day 155.

We used the information described in the previous section regarding derived velocities and structure of RS Oph's extended emission  as initial parameters in {\it Shape}. Particles within {\it Shape}, each having uniform emissivity, are used to provide a simulated density and velocity field and can be either distributed uniformly on the surface or within the volume of the appropriate structure. Since the emission line of interest, \oiii, is very broad and extends beyond the blue cut-off of the \hst\ filter, particularly for the ACS/HRC camera, we also applied a Doppler filter to replicate the finite extent of the \hst\ narrow band filters (in the case of F502N of the ACS, the central wavelength of the filter, width 57\AA\, is 5022\AA\, so the sensitivity to material outside the range $\sim -800 \textrm{ to } +2600$ km s$^{-1}$ is greatly reduced - see below).

This modeling also allows us to determine nebular remnant inclination and relate this to that of the central binary (where an inclination $i$ = 90\degree\ corresponds to the major axis of the bipolar structure or the orbital axis of the binary lying in the plane of the sky). Different authors have suggested that the central binary system has an inclination between 30\degree\ $\le$ $i$ $\le$ 40\degree\ \citep{DK94} or 49\degree\ $\le$ $i$ $\le$ 52\degree\ \citep{BQM09}.

\section{RESULTS AND DISCUSSION}\label{discussion}
\subsection{General Structure}
Figure \ref{fig:hst_obs} suggests that the evolving resolved structure in RS Oph is bipolar and it has been modeled as such by \citet{OBP06} and in \citetalias{BHO07}. Assuming a bipolar model, we then varied the inclination of the system. We also adopted a linear velocity field given by
\begin{equation}
V_{\textrm{exp}} = \frac{3200}{\sin i} \frac{r}{r_{0}} \textrm{ \kms,}
\end{equation}
where $r$ is the distance of a particle from the center of the remnant and $r_0$ is the true semi-major axis. The waist was modeled so that the north-south extent was constrained to 3.60 $\times$ 10$^{15}$ cm in line with that derived in Section \ref{results_hst} above. In {\it Shape} first we introduced the dumbbell structure with $3\times10^6$ particles distributed on the surface (as the image suggested the outer lobes were not centrally filled) and ran the model several times to produce a well sampled model spectrum. The model image was then pixelated to replicate the \hst\ ACS/HRC CCD pixel size of 0.025 arcseconds.

This simple model does not reproduce the observed spectrum. In particular, it gives too broad a line profile, suggesting that lower velocity material must dominate the emission at $t = 155$ days. We therefore introduced an over-density towards the center containing lower velocity material. This was accomplished via an inner hour glass structure with semi-major axis 1.26 $\times$ 10$^{15}$ / $\sin$ $i$ cm (for $d = 1.6$ kpc) and with the particles distributed within the volume as the central structure showed no evidence of limb brightening in the {\em HST} image. It should be noted that the precise geometry of the inner structure may not be as well constrained as that of the outer dumbbell. However, we have for example explored a simple spherical geometry for this inner region, but the fits to the observed spectra were then much poorer in terms of the gross line profile (inverted U-shape) and finer detail (no double central peak). It should also be noted that this inner structure is smaller than the estimated extent of the red giant wind at this epoch. 

\begin{figure}
\plotone{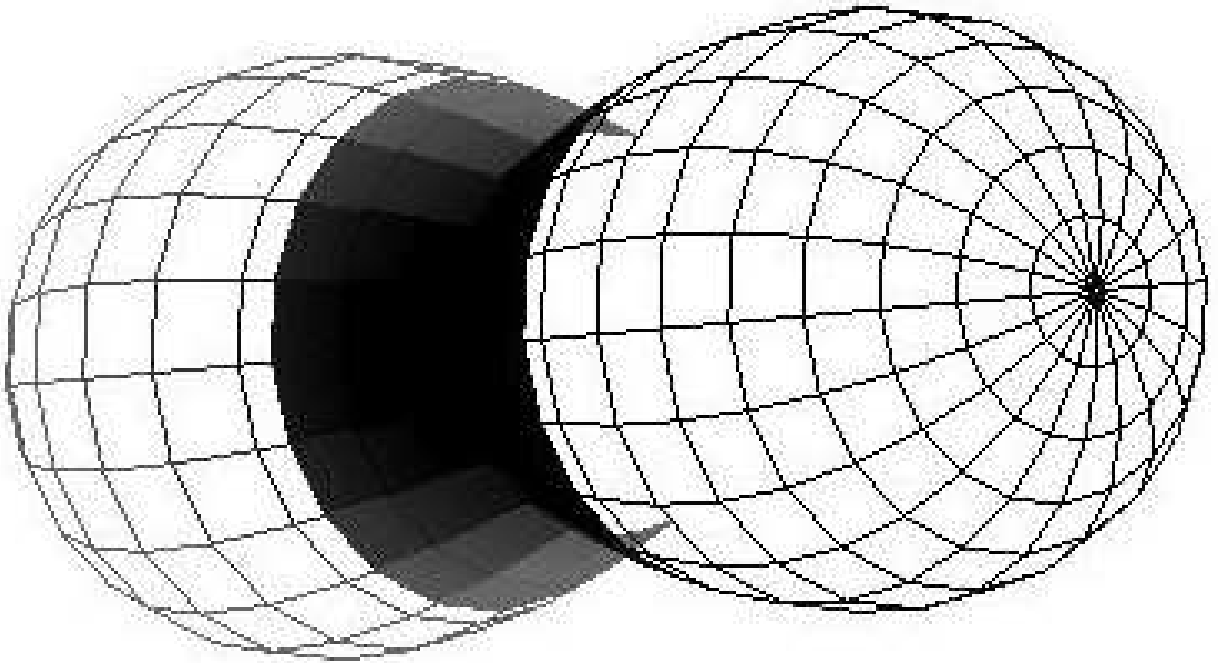}
\caption{Dumbbell (outer) and hour glass (inner) structure that is used to replicate the RS Oph remnant.}
\label{fig:mesh}
\end{figure}

The model system thus contains two structures with the central hour glass having four times more emitting particles than the outer dumbbell structure (Figure \ref{fig:mesh}); this ratio is in approximate agreement with the estimated ratio of flux from the inner region of the image to that in the lobes (factor of 4; likely to be a lower limit for this ratio as some of the flux in the central region will be contributed by the innermost parts of the outer structure).

The existence of a distinctly two component structure is compatible with early infrared interferometry \citep{CNM07,C08}, and radio observations of a central  peak (thermally dominated) and outer lobes (non-thermal; \citealt{TDP89,SRM08,EOB09}). In addition, Vaytet et al. (in preparation) have modeled the early time X-ray spectra of RS Oph and found that they  also required a density enhancement towards the central regions to give a high enough absorbing column.

\subsection{System Orientation}
\begin{figure}
\plotone{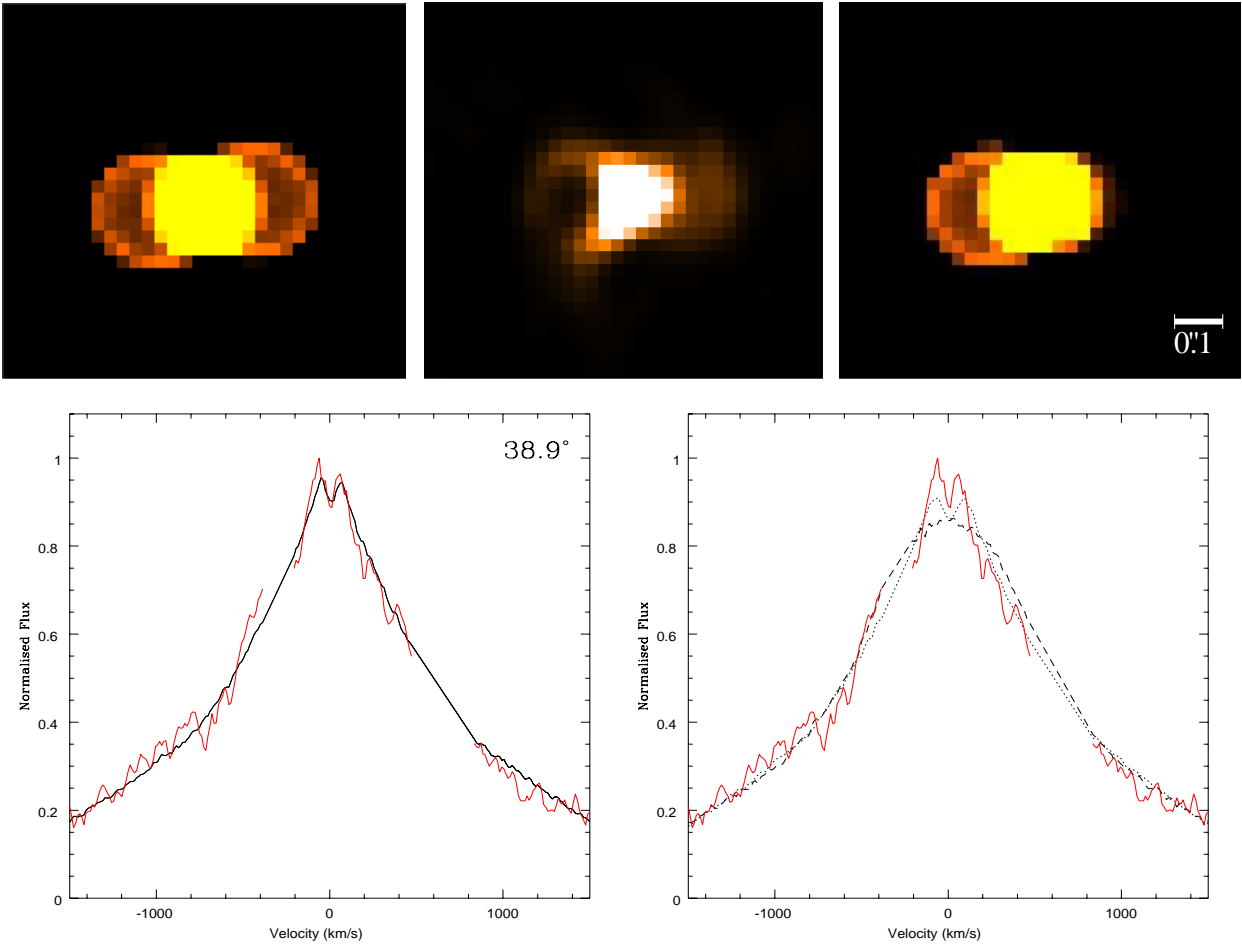}
\caption{{\it Top} - Synthetic image without the Doppler filter applied 
(left), enlarged \hst/ACS image at $t = 155$ days (middle) and synthetic image with 
Doppler filter applied (right). The synthetic images are from a combined 
outer dumbbell and inner hour glass structure inclined at 39 degrees 
to the line of sight and with a position angle of 85\degree. {\it Bottom} - Best fit synthetic spectrum (black; left) is 
overlaid with the observed spectrum (red; the emission lines of \feii\ 
$\lambda$5001.91, \hei\ $\lambda$5015.68\ and \feii\ $\lambda$5018.44\ 
associated with the ionized wind ahead of the forward shock have been removed). Model spectra resulting from inclination angles at the 1$\sigma$ error limits are also shown (right) where the dotted line is for $i$ = 40\degree\ and the dashed line is for $i$ = 29\degree. }
\label{fig:model}
\end{figure}

Figure \ref{fig:model} shows the results of modeling using the outer dumbbell and inner hour glass structures described above. The model image reproduces well the general morphology seen in the {\ HST} image, including that of the west side appearing less prominent than the east. This asymmetry is an observational effect, due to the finite size of the \hst\ filter F502N (as illustrated in Figure \ref{fig:model}; see also Section 4 above) and immediately tells us that the western lobe is approaching the observer. Indeed, any apparent disagreements between the model and observed images can largely be attributed to artefacts of the PSF subtraction procedure. We note that the spectrum shows excess emission on the blueward side (top Figure \ref{fig:evolution}). This could be attributed to \feii\ $\lambda$5001.91, although this line is not observed in spectra at earlier times (Figure \ref{fig:evolution}).

Although radio observations led \citet{OBP06} to model the RS Oph remnant as a bipolar structure, the deduced orientation of the system differs from that found from our results in the optical. \citet{OBP06} modeled the orientation with the east lobe closest to the observer since that is where a second component appears first in their radio observations at early times. However, our observations suggest that the west lobe must be approaching the observer. This is in agreement with results from infrared interferometry at early times and it is suggested that the early emergence of the eastern-most radio lobe may be due to obscuration by a flared disk of circumbinary material lying in the orbital plane \citep{CNM07,C08}. Even at the epoch of the first VLBI observation at $t = 13.8$ days, this material has to be external to the binary and would therefore arise from an anisotropic red giant wind. We also find that the system has a position angle on the sky of 85\degree\ in line with that found from the radio observations of the 1985 outburst by \citet{PDG87}.

\subsection{System Inclination}
{\it Shape} also allows us to determine the inclination of the system to the line of sight. If the nebula's orientation is linked to that of the central binary, and with the knowledge that RS Oph is not an eclipsing system, we calculated the upper limit of $i$ to be used in our exploration of the inclination angle. For an eclipsing binary with an M2 III giant with a radius of 67$_{-16}^{+19}\rsun$ \citep{DS98} and a separation between the giant and WD of $\simeq 1.5$ AU (if for example $M_{WD}+M_{RG} = 2$M$_{\odot}$ and $P_{orb} = 455$ days \cite[see e.g][]{DK94,FJH00,BQM09}) we find that eclipses will occur if $i \ge$ 78$\pm{3}$ degrees. A lower limit for the range over which we explored the model was constrained by velocities being far too high to arise from a nova explosion; e.g. at $i$ = 20\degree, $V_{\rm{exp}} = 9300$ km s$^{-1}$. We therefore ran the model for 20\degree\ $\le i \le$ 81\degree\ and determined each model's respective $\mathcal{X}^2$ fit by comparing the observed and model spectra. 

\begin{figure}
\plotone{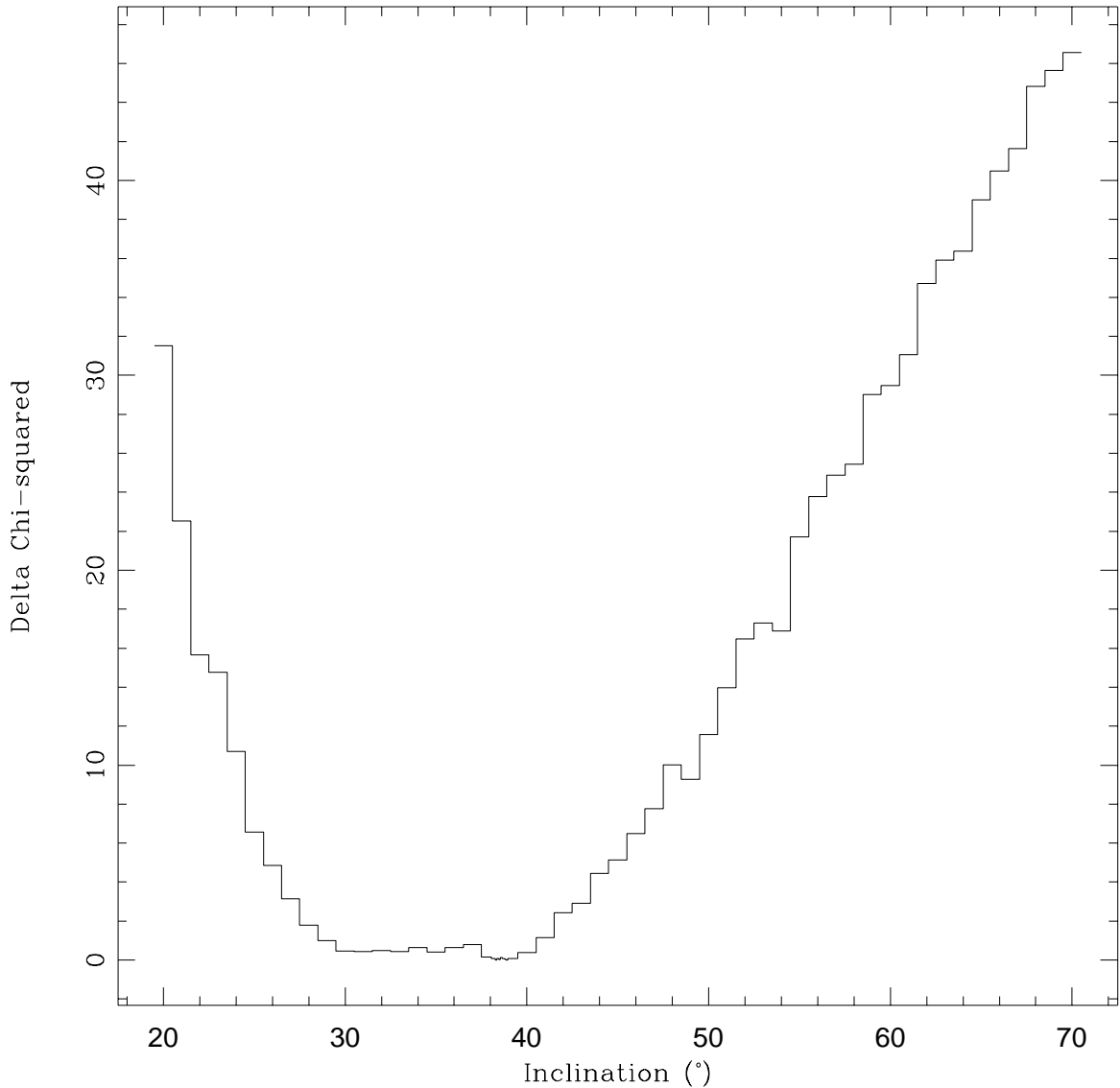}
\caption{$\Delta\mathcal{X}^2$ as a function of inclination angle. This reveals that the best-fit value for the inclination of the system is 39$^{+1}_{-10}$ degrees (1$\sigma$ errors; 3$\sigma$ errors are +8\degree\ and -14\degree). The inclination was varied in 1\degree\ steps and then, when close to the minimum, in 0.1\degree\ steps (note that we explored inclination angles up to 81\degree\ as suggested in section 5.3, but the models continued to provide a very poor fit to the data at such high inclinations).}
\label{fig:chi2}
\end{figure}

We find the best value for the inclination of the remnant to be $i = 39^{+1}_{-10}$ degrees (Figure \ref{fig:chi2}; the errors were determined in the standard way by taking the $\Delta\mathcal{X}^2$ and calculating the 1$\sigma$ level). This is in good agreement with the inclination for the central binary found by \citet{DK94} of 30\degree\ $\le$ $i$ $\le$ 40\degree, though less so with the more recent determination of 49\degree\ $\le$ $i$ $\le$ 52\degree\ by \cite{BQM09}. For illustration, Figure \ref{fig:model} also shows the synthetic spectra derived at the error limits quoted for the inclination where clearly the synthetic line profiles do not reproduce as well the observed profiles as does the best fit value of $i=$ 39\degree. We note that lines of \feii\ $\lambda$5001.91, \hei\ $\lambda$5015.68 and \feii\ $\lambda$5018.44 have been removed from the observed spectrum prior to model fitting, following the procedure outlined in Section \ref{results}, to ensure that our spectral fits are only to the \oiii\ line itself.

\subsection{Second Epoch}
\begin{figure}
\plotone{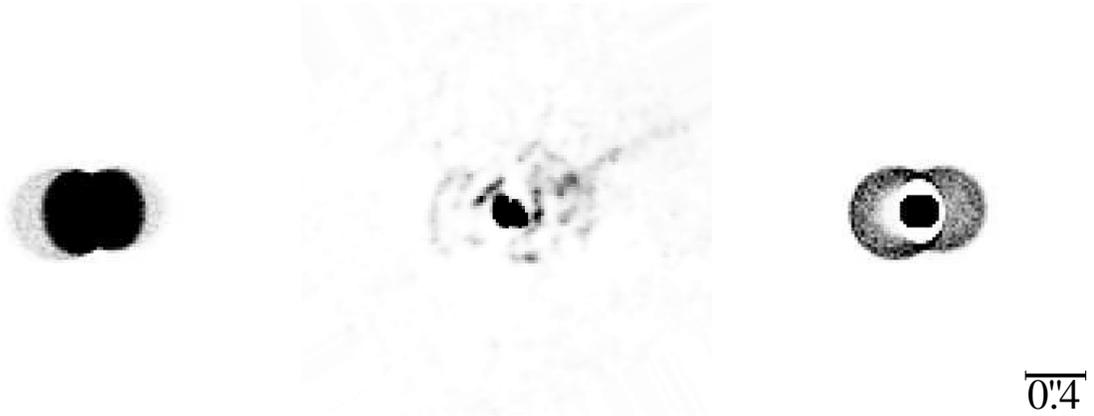}
\caption{{\it Left} - results from a linear expansion of the first epoch model. {\it Middle} -  \hst\/ WFPC2 image at $t = 449$ days. {\it Right} - result from a model in which the outer dumbbell structure was expanded linearly but the central hour glass was kept the same size as in the first epoch. The model images have a Doppler cuts applied at around $-$1600 \kms\ and +2300 \kms\ as appropriate for the WFPC2 filter. North is up and east is to the left.}
\label{fig:model2}
\end{figure}

We also modeled the second epoch \hst\ observation at $t = 449$ days. Using the built-in time evolution option in {\it Shape}, we first explored a simple linear expansion from the first to the second epoch using the results found in Section 4. In this case, the effective wavelength (5012\AA) being much less offset from the rest wavelength of the \oiii\ line itself for the same filter in the ACS means that the asymmetry in the resulting image introduced by the Doppler cuts is not as marked as it was for the first epoch (ACS) image. Figure \ref{fig:model2} shows the results. 

Little resemblance is evident between the observed and modeled images in this case. This is an important result because it tells us that the second epoch is not just a linear expansion of the first, at least in the central regions. We then allowed the dumbbell structure to expand linearly but the inner hour glass was kept the same size as in the first epoch (right hand image in Figure \ref{fig:model2}). We may understand the linear expansion of the outer structure if, as we calculated above, it had already cleared the pre-existing red giant wind by the first epoch. Alternatively, the apparent constant expansion rate may be consistent with powering by narrow jets as proposed by \citet{SRM08}. The apparent constancy of the inferred size of the innermost hour glass structure is more difficult to understand. However, we should be aware of over-interpretation of the poorer image from the second epoch, plus the lack of contemporaneous optical spectroscopy to aid our modeling. Although the model reproduces some other features of the second epoch observations, for example the gap in the east lobe due to the applied Doppler filter, further investigation is required.

\section{SUMMARY}\label{future}
We have reported combined \hst\ imaging, ground-based spectroscopy and detailed modeling of the RS Oph nebular remnant following the 2006 outburst. We find that:
\begin{enumerate}
\item The remnant of RS Oph can be understood as a bipolar structure with two co-aligned components. In order to satisfactorily model both the image and spectra from day 155, we require the presence of a high density (low expansion velocity) central region and a more extended (higher expansion velocity) less dense region in the form of two lobes. This morphology is in agreement with that inferred from observations in the radio and at X-ray wavelengths, plus infrared interferometry at early times.
\item The asymmetry observed in the first epoch \hst\ image is due to the finite width of the \hst\ filter. This in turn implies that the west lobe is approaching the observer, in agreement with the results of infrared interferometry at early epochs.
\item We are able to determine the inclination of the remnant as 39$^{+1}_{-10}$ degrees. This is comparable to estimates of the inclination of the central binary of \citet{DK94}, but lies outside the range of $i$ from recent estimates by \citet{BQM09}.
\item Modeling of the second epoch \hst\ image implies that the outer dumbbell structure underwent linear expansion; however, there is more evidence of deceleration for the central (hour glass) region.
\end{enumerate}

\acknowledgments
The authors are very grateful for the \hst\ Director for provision of Discretionary Time. We are also grateful to M. Shara for helpful discussions regarding our \hst\ program. The authors thank C. Simpson for advice regarding the spectroscopic data, W. Steffen and N. Koning for valuable discussions on the use of {\it Shape} and adding special features to the code and P. A. James for reading an initial draft. VARMR is funded by an STFC Studentship. SS acknowledges partial support from NASA and NSF grants to ASU. JB, VHCh and MGR gratefully acknowledge financial support from DGAPA-UNAM projects 108406, 108506, 116908 and 102607-3, as well as CONACyT grants 54480-F, 43121 and 82066. We thank an anonymous referee for valuable comments on the original manuscript.

{\it Facilities:} \facility{HST (ACS/HRC, WFPC2)}, \facility{Observatorio Astron\'{o}mico Nacional en San Pedro M\'{a}rtir}, \facility{Observatorio Astrof\'isico Guillermo Haro at Cananea}

\clearpage

\begin{deluxetable}{lccc}
\tablecolumns{4}
\tablewidth{0pt}
\tablecaption{Log of optical spectral observations.\label{tb:spectra_log}}
\tablehead{
Date (days & \multirow{2}{*}{Instrument} & Total Exposure & \multirow{2}{*}{R} \\
after outburst) & & time (s)
}
\startdata
2006 February 22 (10) & Echelle & 430 & 16000 \\
2006 February 23 (11) & Echelle & 425 & 16000 \\
2006 February 24 (12) & Echelle & 435 & 16000 \\
2006 February 25 (13) & Echelle & 435 & 16000 \\
2006 March 14 (30) & Echelle & 600 & 16000 \\
2006 March 15 (31) & Boller \& Chivens & \phn90 & 2355.2 \\
2006 April 17\tablenotemark{\star} (64) & Boller \& Chivens & 150 & 1668.6 \\
2006 April 18\tablenotemark{\star} (65) & Boller \& Chivens & 140 & 3149.6 \\
2006 May 29 (83) & Boller \& Chivens & \phn40 & 1525.9 \\
2006 June 18 (126) & Echelle & 1830 & 16000 \\
2006 June 19 (127) & Echelle & 990 & 16000 \\
2006 July 30 (168) & Echelle & 420 & 16000 \\
\enddata
\tablenotetext{\star}{Observations taken at the Observatorio Astrof\'isico Guillermo Haro, otherwise from Observatorio Astron\'omico Nacional en San Pedro M\'artir.}
\end{deluxetable}

\end{document}